\documentclass [12pt,twoside]{article}           
\usepackage[left,modulo]{lineno}
\usepackage{setspace}

\countdef\arxiv=201

\ifnum\arxiv=0 \input cpreb.tex \fi

\ifnum\arxiv=1

\usepackage{epsfig,times,lscape}
\usepackage[usenames]{color}

    \ifnum\count102=1

    \fi

    \ifnum\count102=2
\fi

\newcommand{\nc}{\newcommand}

     \ifnum\count101=1
\nc{\qI}[1]{\section{{#1}}}
\nc{\qA}[1]{\subsection{{#1}}}
\nc{\qun}[1]{\subsubsection{{#1}}}
\nc{\qa}[1]{\paragraph{{#1}}}

\def\qpar{\vskip 2mm plus 0.2mm minus 0.2mm}
\def\qL{\hfill \break}
     \fi 

      \ifnum\count101=2
 \nc{\qI}[1]{\parindent=0mm \vskip 8mm 
{\centerline{\LARGE \color{red}#1}}\vskip 3mm}
\nc{\qA}[1]{\vskip 2.5mm \noindent 
{{\bf\large\color{blue}  #1}} \vskip 1mm \parindent=0mm}
 \nc{\qun}[1]{\vskip 1mm \noindent {\sl #1 }\quad }

\def\qL{\hfill \break}
\def\qpar{\vskip 2mm plus 0.2mm minus 0.2mm}

      \fi

\nc{\qfoot}[1]{\footnote{{#1}}}

      \ifnum\count101=1
\def\qbu{\hfill \par \hskip 6mm $ \bullet $ \hskip 2mm}
\def\qee#1{\hfill \par \hskip 6mm (#1) \hskip 2 mm}
      \fi
      \ifnum\count101=2
\def\qbu{\hfill \par \hskip 4mm $ \bullet $ \hskip 2mm}
\def\qee#1{\hfill \par \hskip 4mm (#1) \hskip 2 mm}
      \fi

\def\qparr{ \vskip 1.0mm plus 0.2mm minus 0.2mm \hangindent=10mm
\hangafter=1}

     \ifnum\count101=1 
 \def\qdec#1{\parindent=0mm\par {\leftskip=2cm {#1} \par}}
     \fi
     \ifnum\count101=2
  \def\qdec#1{\parindent=0mm \par {\leftskip=1cm {#1} \par}}
  
  \def\qcitb#1{\noindent \hbox to 102mm{\hfill \small #1} \vskip 1mm}
      \fi

 \def\qpages#1{\count102=0{\loop\advance\count102 by 1
 \null \vfill\eject \ifnum\count102<#1 \repeat}}

\def\qv{\vskip 0.1mm plus 0.05mm minus 0.05mm}
\def\qhu{\hskip 0.6mm}
\def\qhv{\hskip 3mm}

\def\qhw{\hskip 1.5mm}
\def\qleg#1#2#3{\noindent {\bf \small #1\qhw}{\small #2\qhw}{\it \small #3}\qv }
\fi

\begin{document}
\thispagestyle{empty}



\markboth{{\sl \hfill  \hfill \protect\phantom{3}}}
        {{\protect\phantom{3}\sl \hfill  \hfill}}

\color{yellow} 
\hrule height 10mm depth 10mm width 170mm 
\color{black}

 \vskip -12mm   
%

\centerline{\bf \Large Exploration of the strength of
family links}
\vskip 3mm
 \vskip 10mm

\centerline{\large 
Peter Richmond$ ^1 $ and Bertrand M. Roehner$ ^2 $
}

\vskip 10mm
\large

%
{\bf Abstract}\qL
Ever since the studies of Louis-Adolphe Bertillon 
in the late 19th century
it has been known that marital status and number of children
markedly affect death and suicide rates.
This led in 1898 Emile Durkheim to conjecture a
connection between social isolation (especially
at family level) and suicide.
However, further progress was long hampered
by the limited statistical data
available from death certificates.
Recently, it was shown by the present authors
that disability data from census records
can be used as 
a reliable substitute for mortality data. 
This opens a new route to investigations
of family ties because census information goes much beyond
the limited data reported on death certificates.
It is shown that the disability rate of adults
decreases when they have more family links.
More precisely, the reduction of the parents'
disability brought about
by the presence of a child reveals that the strength
of ties between parents and child is highest
in the first year after birth and then weakens steadily
as the child ages.
It will also be seen that the strength of the
bond between husband and wife is highest when they are
of same age and decreases fairly steadily when the age gap 
increases.  
                              
\vskip 10mm
\centerline{\it \small Provisional. Version of 14 July 2017. 
Comments are welcome.}
\vskip 5mm

{\small Key-words:  disability, death,  family, children
parents, age}

\vskip 5mm

{\normalsize
1: School of Physics, Trinity College Dublin, Ireland.
Email: peter\_richmond@ymail.com \qL
2: Institute for Theoretical and High Energy Physics (LPTHE),
University Pierre and Marie Curie, Paris, France. 
Email: roehner@lpthe.jussieu.fr
}

\vfill\eject

\qI{Introduction}

\qA{Influence of family links on death rates}

This paper is a step forward in a story that started in 
the second half of the 19th century. At that time it was
realized that in all countries where data were available
the age-specific death rate of married people was
2 or 3 times lower than
for non-married (i.e. single, widowed or divorced) people. 
This became known as the Farr-Bertillon law%
\qfoot{William Farr (1807--1883) and
Louis-Adolphe Bertillon (1821--1883) were
epidemiologists regarded as two of the founders of medical
statistics.}%
.
Subsequently, Louis-Adolphe Bertillon (1872,1879) extended his
1872 study to include 
the influence of the number of children on the suicide rate%
\qfoot{Bertillon's results are summarized in Richmond et al.
(2016a).}%
.

\qA{Durkheim's conjecture}

From this work emerged the conjecture that greater
social isolation due to a lack of short-range 
husband-wife or parent-children
interactions produced higher death rates.
This became one of the strongest arguments in favor of  Durkheim's
thesis that the underlying cause of suicide is social isolation.
Later on it was shown (Gove 1973, Richmond et al. 2016a)
that marital status is a significant
factor in age-specific death rates {\it separately} for all major
causes of mortality, e.g. heart, pulmonary, cerebrovascular,
cancer diseases.
\qpar

Incidentally, this observation provides an
answer to the objection which is sometimes raised that
``it is difficult to distinguish the beneficial impact of marital status
on health from the confounding effect of selection into the married
state'' (Christensen et al. 2011). How
can a marriage-based selection affect the death rate due 
(for instance) to
cerebrovascular attacks which occur several decades later?
The conclusion that marriage-based selection plays
a very limited role
is also confirmed by twin studies (Osler et al. 2008).
Another misplaced conception 
is to say that ``marriage has a positive effect on
health by altering preferences for risky behavior''.
(Santerre et al. 2013, p. 50). Even if correct in some
cases, this explanation can account 
only for mortality by accidents which is a very limited
aspect of the question.

\qA{Biodemographical experiments in animal populations}

In order to test Durkheim's thesis in a broader 
and more systematic way,
it is of course tempting to make observations on
animal populations. This idea was tried in
an experiment done with ants and fruit flies at the 
``South China Agricultural University'' 
It lasted over two years from 2012 to
2014 (Wang et al. 2016). Individuals were extracted from
the rest of their population in two ways.
\qbu Single individuals
\qbu Groups of 10 individuals.
\qpar
In both cases the extracted insects were given 
appropriate and identical living conditions
in terms of temperature, hygrometry, light and food.
Two observations were of particular significance.
\qbu Firstly, the two or three days immediately 
after extraction were
marked by a surge in death rates.
\qbu Secondly,
in the course of time the average death rates
were higher among the ``singles''
than in the groups of 10. This was observed for the
ants as well as the fruit flies.

\qA{Need of more detailed statistical sources}

However, for a better understanding of the connection
between social isolation and death rates there was a major
obstacle in the sense that statistical information
about deceased people was limited to the data recorded
on death certificates, i.e. 
age and marital status of the deceased
and cause of death (immediate as well
as underlying cause) as determined by a physician.
\qpar

Many other data would have be of great interest
from the perspective of social isolation,
particularly data
about the children of deceased persons (e.g.
number and age) for, apart from the husband-wife
interaction, the parent-child interaction
is certainly the most important social link.
Because of this lack of data
the conjecture of a connection between
mortality and social isolation could not be tested further.

\qA{Work disability as a substitute for mortality data}

Progress became possible when it was
found that disability data from census records could be
used as a reliable substitute for mortality rates
(Richmond et al. 2017).
\qpar
More specifically, the
following question was asked in the US censuses 
of 1980 and 1990%
\qfoot{Similar questions were asked in other years
but which did not have the same accuracy. 
For instance, in all ``American Community Surveys'' (ACS) 
from 2000 to 2007 the following question was asked:
\qdec{``Because of a physical, mental, or emotional condition lasting 6
months or more, does this person have any difficulty in working at a
job or business?''.} 
Clearly there is more subjectivity in assessing
a difficulty than an impossibility.}%
:
\qdec{``Does this person have a physical, mental, or other health
condition that has lasted for 6 or more months and which
prevents this person from working at a job?''}
\qpar

Several tests (Richmond et al. 2017) all
showed that the proportion of persons with complete work disability
is strongly correlated with the annual mortality rate.
For instance the fraction with disability increases exponentially
with age (see Fig. 1,2,3) in a way which parallels
the Gompertz law which describes the increase of the death rate.
\qpar

The paper proceeds as follows.
\qbu First, we briefly explain the procedure
(more detailed explanations can be found in appendix A)
and we give an overview in the form of three graphs
which illustrate two possible presentations of the 
results.
\qbu Then we investigate in more detail the effect of the age 
of the child on the disability of the parents.
\qbu Finally, we investigate the effect on disability of an age gap
between husband and wife.

\qI{Overview of age-specific disabilities}

\qA{Procedure}

The procedure is fairly straightforward. Nevertheless,
for the benefit of readers who would like to
repeat the calculations or do similar investigations,
the practical details of the procedure are explained
in Appendix A. 

\qA{Introduction of two metrics for the analysis of disability data}

The key-variable is the work disability. We limit ourselves to
two cases: no disability versus disability which prevents the person
from working; in other words, we discard intermediate cases
in which working is made more difficult but is still possible.
\qpar
The two metrics that we shall use parallels the metrics commonly
used for the analysis of age-specific death rates.
\qbu For any group, $ A $, defined by a specific
marital situation and a given age group
the {\it disability likelihood}
is obtained by dividing the number of
persons in $ A $ who have a disability
by the total population of $ A $. We denote this probability
by $ P1 $. It was shown elsewhere (Richmond et al. 2017) that
$ P1 $ is closely related to the death rate%
\qfoot{We use the term ``rate'' because it is commonly used
but it should be recalled that in fact the death rate
represents a probability,
namely the likelihood of dying in the age interval under
consideration.}
\qbu Similarly to the death rate ratios and for the same reason,
namely to get rid of the exponential growth
one can define disability ratios.
\qpar

The following subsections review these variables for
several situations of interest.

\qA{Social status versus interaction between spouses}

With respect to marital situations a question comes to mind
immediately which is of both practical and conceptual interest:
for married persons is it the {\it social status} which matters 
(in terms of disability) or the actual {\it interaction} 
between spouses?
%
\begin{figure}[htb]
\centerline{\psfig{width=12cm,figure=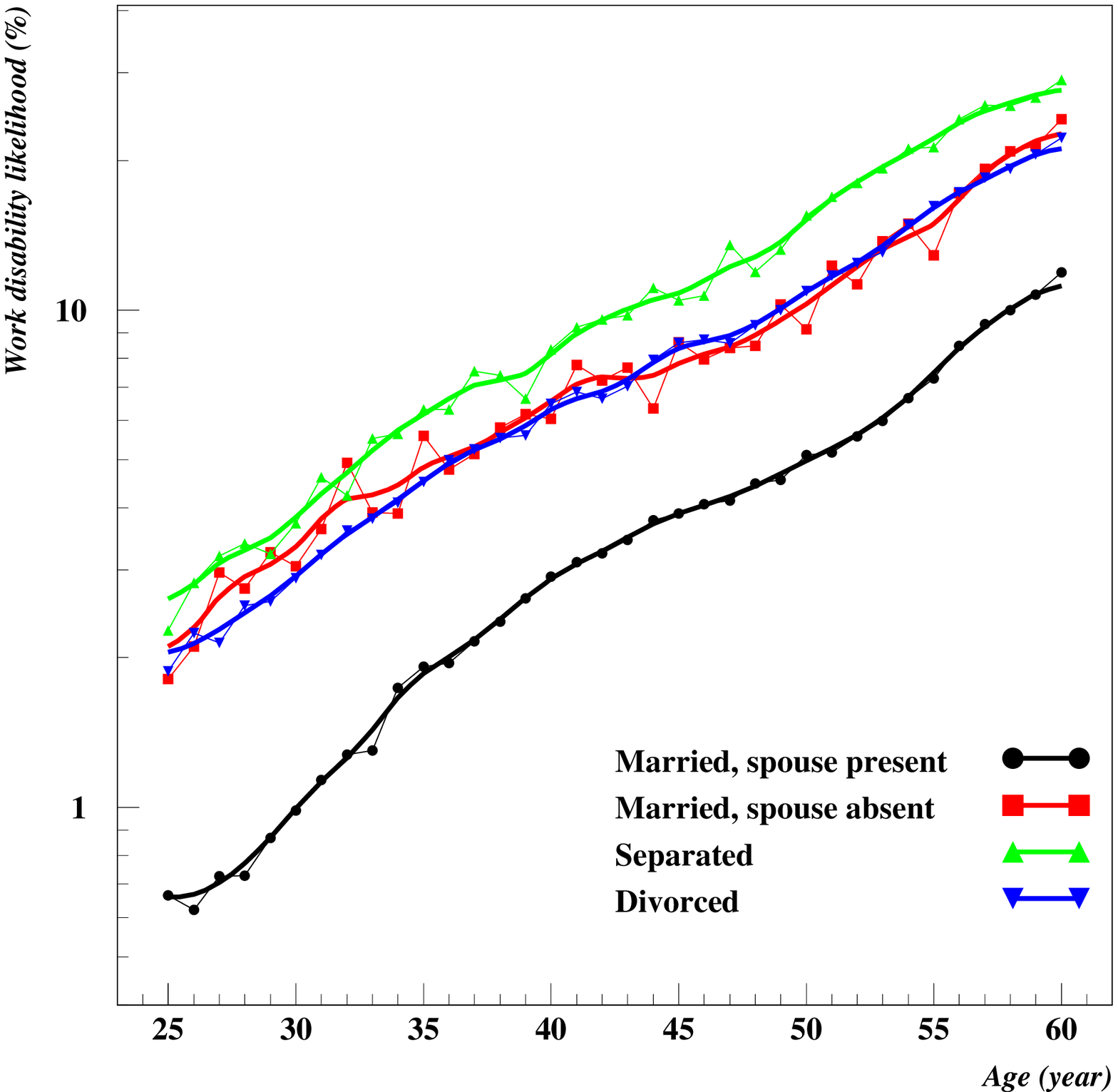}}
\qleg{Fig.\qhu 1\qhv Likelihood of disability
in various marital situations 
(for males and females together in 1990).}
{In view of their disability levels
the four curves fall into two classes: ``married with
spouse present'' on the one had and on the other hand
there are the three curves ``married with spouse
absent''+``separated''+``divorced''.
In other words, it is the real presence of the spouse 
which matters, rather than the social status of
being married.}
{Source: 5\% sample of the US census of 1990 downloaded from the 
IPUMS database (Ruggles et al. 2017).}
\end{figure}
%
What we mean is the following. There is a pressure to get married
exercised on young adults by families as well as by society (more
so in former times than nowadays) which can lead to a
substantial amount of stress. Getting married is a way to remove
that pressure and feel better. In this status explanation
the interaction between spouses is either downplayed or
ignored.
\qpar
Fig. 1 allows a clearer insight into this question.
Among the groups considered, there
are two, namely ``married, spouse absent'' and ``separated''
which provide the status of being married
without however any real presence of a
spouse. It can be seen that the disability likelihood
of these groups is much closer to the case of divorced persons
than to the
case of married persons with spouse present. 
So, this graph strongly suggests that  
the actual interaction between spouses matters much
more than the social status alone.
\qpar
Incidentally, one may find surprising
that the curve of ``separated'' is slightly
above ``divorced'' for the reason that ``divorced''
seems a more serious and definitive situation. 
A possible explanation
may present itself when one refers to what ``separated''
really means in the census. It applies (i) to
persons who have parted by themselves because of marital discord 
(ii) to couples  who have already a legal separation prior
to an upcoming divorce. We see that both cases
refer to the early stage of a separation. It is known
(see Richmond et al. 2016b) that for events such as divorce,
widowhood there is a death rate spike 
which lasts about 6 months and surges above the stationary
death rate of the initial (married) situation. Obviously,
the weight of such spikes will be more diluted in
longer time intervals, that is to say for the ``divorced''
category.

\qA{Age-specific disability in various family situations}

Fig. 2 shows curves of disability likelihood
for various family types.
Three observations can be made.

%
\begin{figure}[htb]
\centerline{\psfig{width=12cm,figure=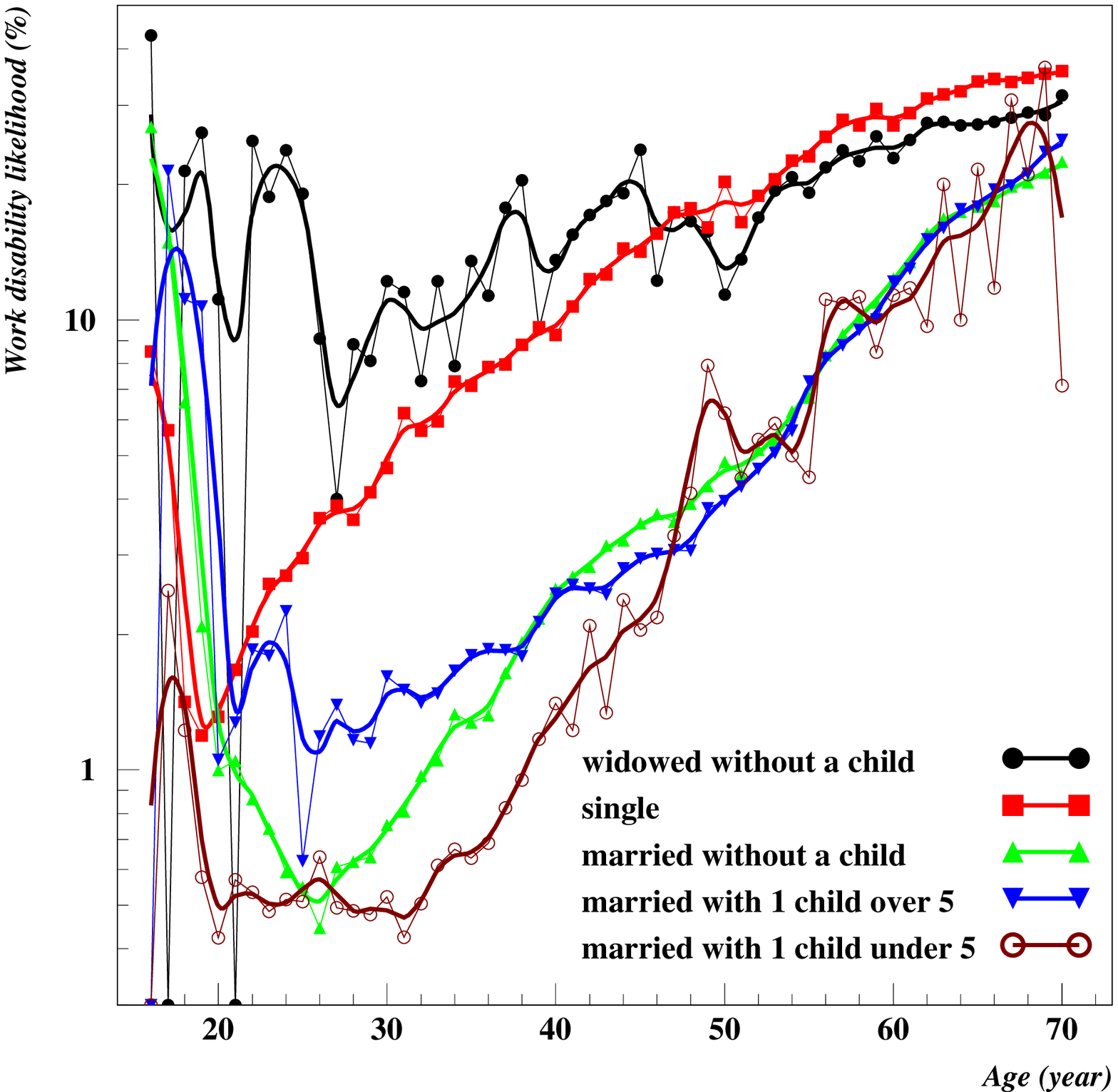}}
\qleg{Fig.\qhu 2\qhv Likelihood of disability
in various family configurations (for males in 1990).}
{The amplitude of the fluctuations (as shown by the
yearly data points) is mostly determined by the number of
persons in each age group. The well-known ``young widower 
effect'' has an amplitude of more than 10.
While for some reason not yet well understood 
this is specific to males, the other results display the same
pattern for men and for women. Moreover the fact that 
one gets a very similar graph for the census data of 1980
shows that the observed effects are fairly stable.}
{Source: 5\% sample of the US census of 1980 and 1990;
downloaded from the 
IPUMS database (Ruggles et al. 2017).}
\end{figure}
%
\qbu All curves display an upward exponential trend 
(albeit with different exponents)
which
is in line with Gompertz's law which says that
death rates increase exponentially with age.
\qbu Although the curves display sizable differences 
particularly in mid-age, they all converge toward
a same limit in old age. Here we restricted 
the analysis to the age interval $ (16,70) $. 
In subsequent graphs it will be seen
that this convergence continues in the age interval $ (70,90) $.
It has a simple interpretation; it means that 
in old age purely biological factors of aging develop
and become predominant. As a result, the differences 
due to marital circumstances become imperceptible.
\qbu Finally, what is from our perspective
the most important observation is the fact that for
any age-group the disability likelihood decreases
as the number of family bonds increases. From widowers
or unmarried persons to married persons with a child 
under 5 years the likelihood is divided by a factor 
between 5 and 20.

\qA{Time constant of the transition from ``single'' to ``married''}

The census of 1880 of which IPUMS gives a 10\% sample offers a
rare opportunity to explore the transition from the state of
being single to the state of being married. 
%
\begin{figure}[htb]
\centerline{\psfig{width=12cm,figure=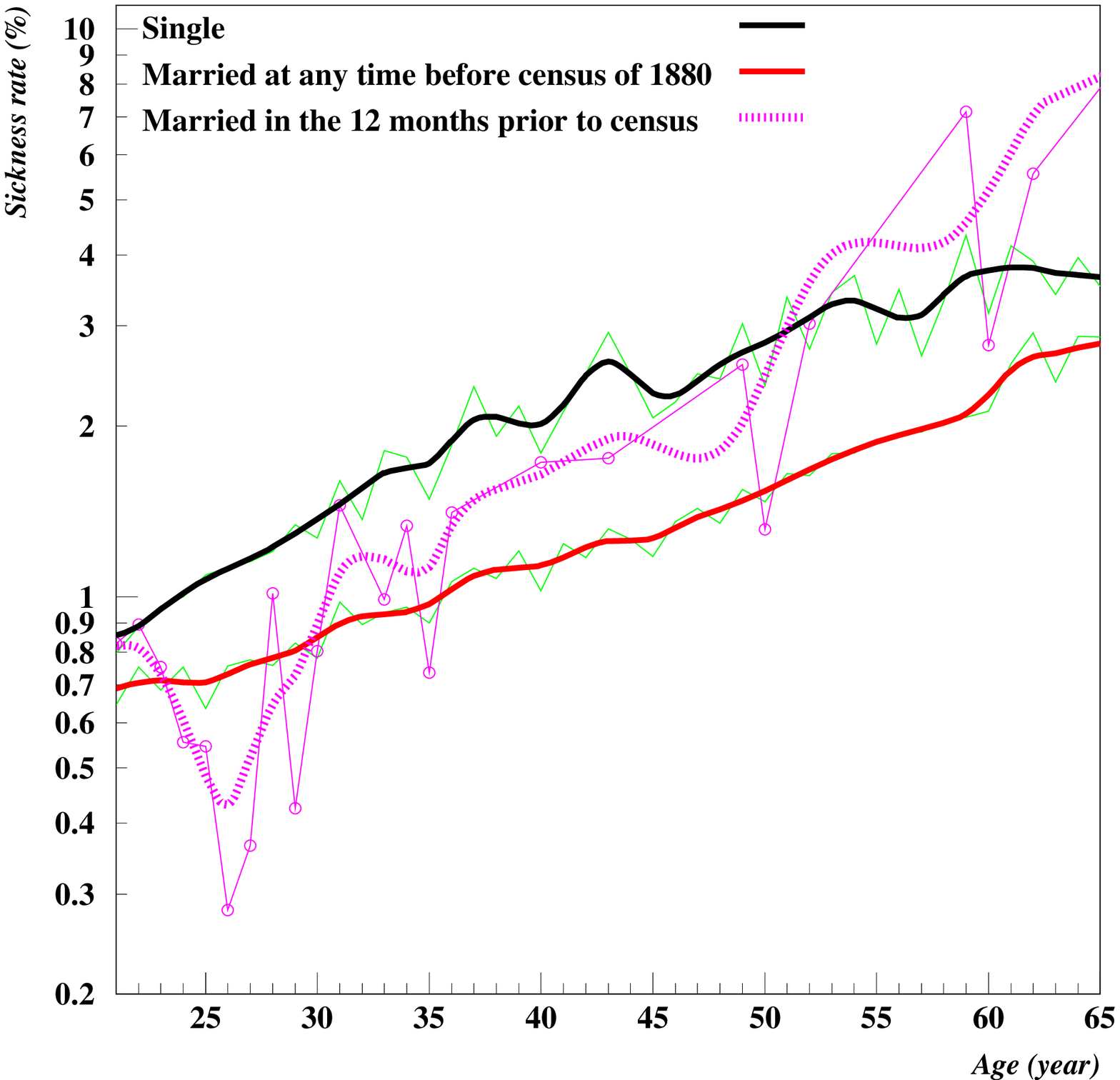}}
\qleg{Fig.\qhu 3\qhv Sickness rate in the transition
from single to married state.}
{The sickness rate gives the proportion of sick persons
at the moment of the census. The dotted curve
corresponds to an average delay of 6 months between
marriage and census time. At that moment the
sickness rate is still intermediate between the
rates for single and married. As shown by the
amplitude of the fluctuations, the confidence interval
becomes fairly large in old age when there are
few newly married persons.}
{Source: 10\% sample of the US census of 1880;
downloaded from the 
IPUMS database (Ruggles et al. 2017).}
\end{figure}
%

This census tells us whether or not 
the respondents got married in the
12 months which preceded the taking of the census.
Whilst in a legal sense getting married
is instantaneous, the physical
and mental state does not change immediately but with a 
time constant that we denote by $ \theta $. 
Is $ \theta $ of the order of a few weeks, a few months
or a few years?
\qpar
If one assumes a
uniform distribution of marriage dates, the 
dotted curve of Fig. 3 corresponds to an average
delay of 6 months between marriage and census time.
The fact that this curve is still only mid-way between
the curves for ``single'' and ``married'' shows that $ \theta $
is of the order of several months. More detailed data
would be needed to give a closer estimate.

\qA{Disability ratio}

The likelihood curves in Fig. 2 give an overall
picture but they are not well suited for 
accurate measurements. In studying the effect
of marital status on death rates it is appropriate
to define a reference case $ C_r $ (for instance the
situation of being married) and then, for any
other marital situation, to compute the ratio with
respect to $ C_r $. This is the procedure
followed by the present authors in Richmond et al. (2016a).
By discarding the exponential growth
trend which is common to all cases, this procedure allows
higher accuracy. 
\qpar

%
\begin{figure}[htb]
\centerline{\psfig{width=16cm,figure=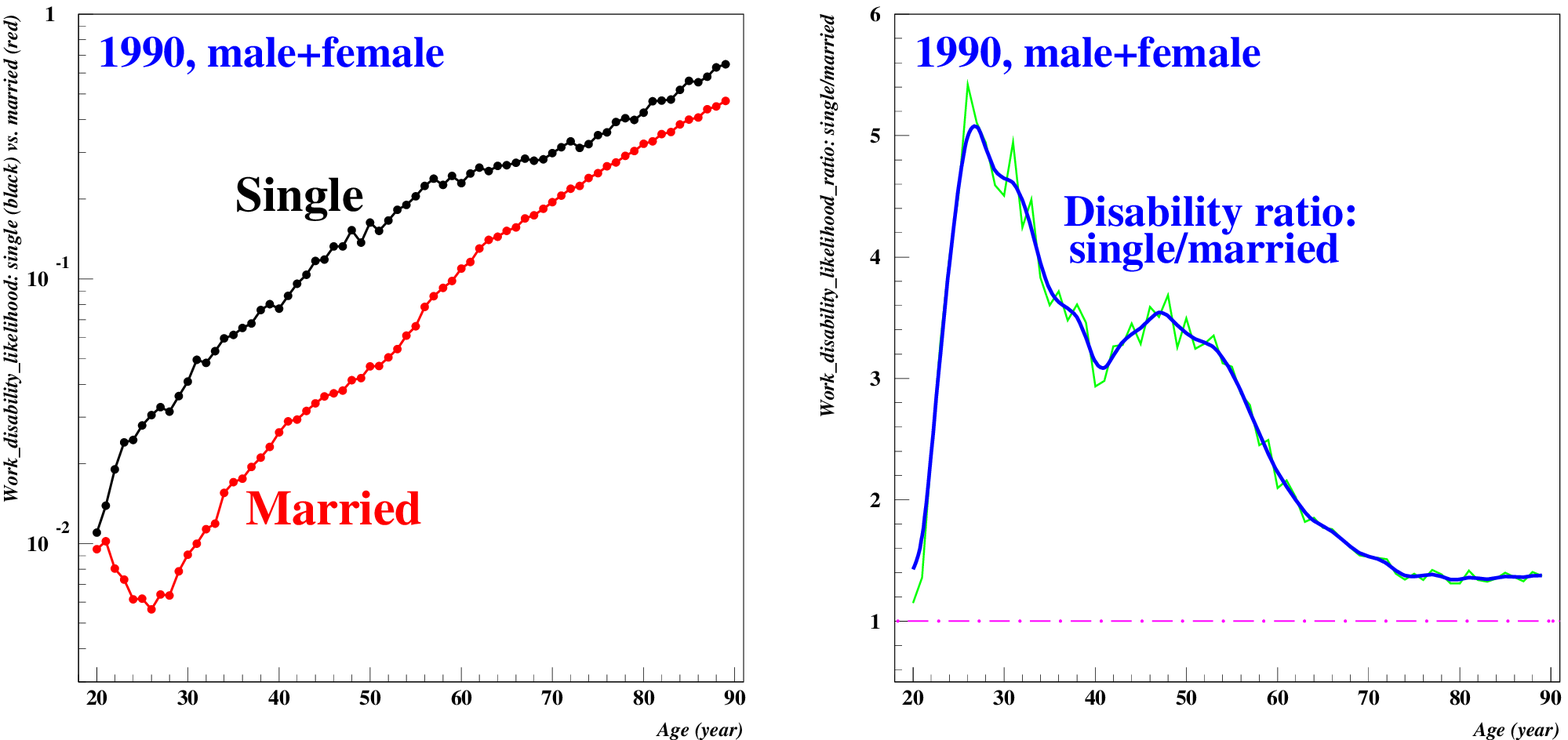}}
\qleg{Fig.\qhu 4a,b\qhv Disability for
single and married persons (in both cases there is nobody
else in the household).}
{{\bf (a)} Disability likelihood for single and married persons.
In the $ (20,32) $ age interval, it is remarkable that 
the benefit of being
married is strong enough to overcome the upward trend
of disability with age.
{\bf (b)} Disability ratio $ R= $single/married.}
{Source: 5\% sample of the US census of 1990 
downloaded from the 
IPUMS database (Ruggles et al. 2017).}
\end{figure}
%

One drawback, however, is that
with this representation one is limited to pairs of variables.
In order to cover the 5 curves of Fig. 2 one 
would need to consider 10 pairs.
Another drawback is the fact that
when the ratio displays a peak one
does not know if it is due to a peak in the
numerator or a trough in the denominator.
\qpar
Notwithstanding these limitations, the disability ratio
will provide accurate assessments of two important
features: 
\qbu Is the ratio larger or lower than one?
\qbu What is the maximum amplitude of the ratio?
\qpar

In Fig. 4a,b the procedure is illustrated for the
pair (single,married).
In Fig. 4b we see that $ R $ is
above 1 for all ages. In other words, 
with respect to non-married persons, the marital bond
provides a protection against disabilities. This effect is quite
considerable in the age range between 25 and 35.

\qI{Connection between disability and social isolation}

\qA{Why a disability ratio larger than one implies some form
of interaction}

In order to get an intuitive perception it is useful 
to consider two simplified cases. They are simplified
in the sense that we assume that the disability ratio $ R $ 
is the same for all ages.
\qbu First, suppose that $ R=1 $. This means that from
the perspective of
personal health, it does not make any difference whether the 
persons are single or married. Such an outcome is consistent
with a complete lack of interaction between spouses. In this
case each spouse would be just like a non-married person.
\qbu Now, suppose that $ R = 5 $. Unless, marriage has pre-selected
the spouses in some way%
\qfoot{An explanation already discussed above 
and which is furthermore
contradicted by the fact that the ratio shows strong variations
with age.}
the state of being married must have a beneficial effect on the spouses.
We have shown above that this is not merely a ``status effect''
but is conditioned by the actual presence of the spouse.

\qpar
Clearly
the interaction between spouses can take many forms,
some of which may have no health effects. For instance,
spouses may cooperate in running
a business together; such a cooperation certainly implies
a form of interaction. 
That specific interaction would have an effect
on {\it economic} variables. 
Thus, if the average income ($ i_2 $)
of couples is higher than twice the income  
($ i_1  $) of single persons involved in the same activity,
then the excess-income $ e=i_2-2i_1 $ may be used
as a measure of the strength and usefulness
of such a business interaction. 
\qpar
Economic cooperation, as considered above, is easier to
conceive intuitively than an interaction resulting in
health effects.
At this point one can only
say that there {\it must} be some form of interaction
that is beneficial to both individuals. 

\qA{Connection between social isolation and high
suicide rates}

The connection between high suicide rates and social
isolation has been documented in many situations
(Roehner 2007, section entitled ``Effect of social
isolation on suicide'' p. 209-220). Let us recall
some of them.
\qee{1} The social disruption of former links 
experienced by immigrants when they leave their country of origin
for a new place results in a multiplication
of suicide rates by a factor of 1.4.
\qee{2} The suicide rate of prison inmates is between 2 and 5 times
higher (depending on the specific type of prison)
than the rate for the general population of same age and gender.
\qee{3} The suicide rates of prisoners in solitary confinement,
in remand centers or in lockup (i.e. immediately after arrest)
are at least three times higher than that of the 
general prison population. 
\qee{4} According to the ``Schizophrenia Society of Ontario'',
the suicide rate of schi\-zo\-phre\-nics is about
10 times higher than that of the general population.
\qpar
Most of the cases mentioned above concern young or middle-aged
persons which means that their death rates for other causes
than suicide are rather low. In other words,
for such people it is not only the suicide rate which is
inflated but also the ``all causes'' death rate.
\qpar

What is the mathematical relationship between
disability and bond strength? \qL
As they vary in opposite direction
one can posit a relationship between
bond strength and the inverse of the disability.
For the moment as we limit ourselves to {\it relative}
estimates of bond strength, we cannot 
state (and in fact do not need to know)
what is the precise mathematical form
of the relationship%
\qfoot{The relationship could be something of the form: 
$ b\sim 1/d^{\alpha} $, where $ b $ is the bond strength,
$ d $ the disability likelihood and $ \alpha $ a positive
parameter.}%
.

\qI{Impact of the presence of children according to their age}

\qA{Presence of a child of any age}
%
\begin{figure}[htb]
\centerline{\psfig{width=16cm,figure=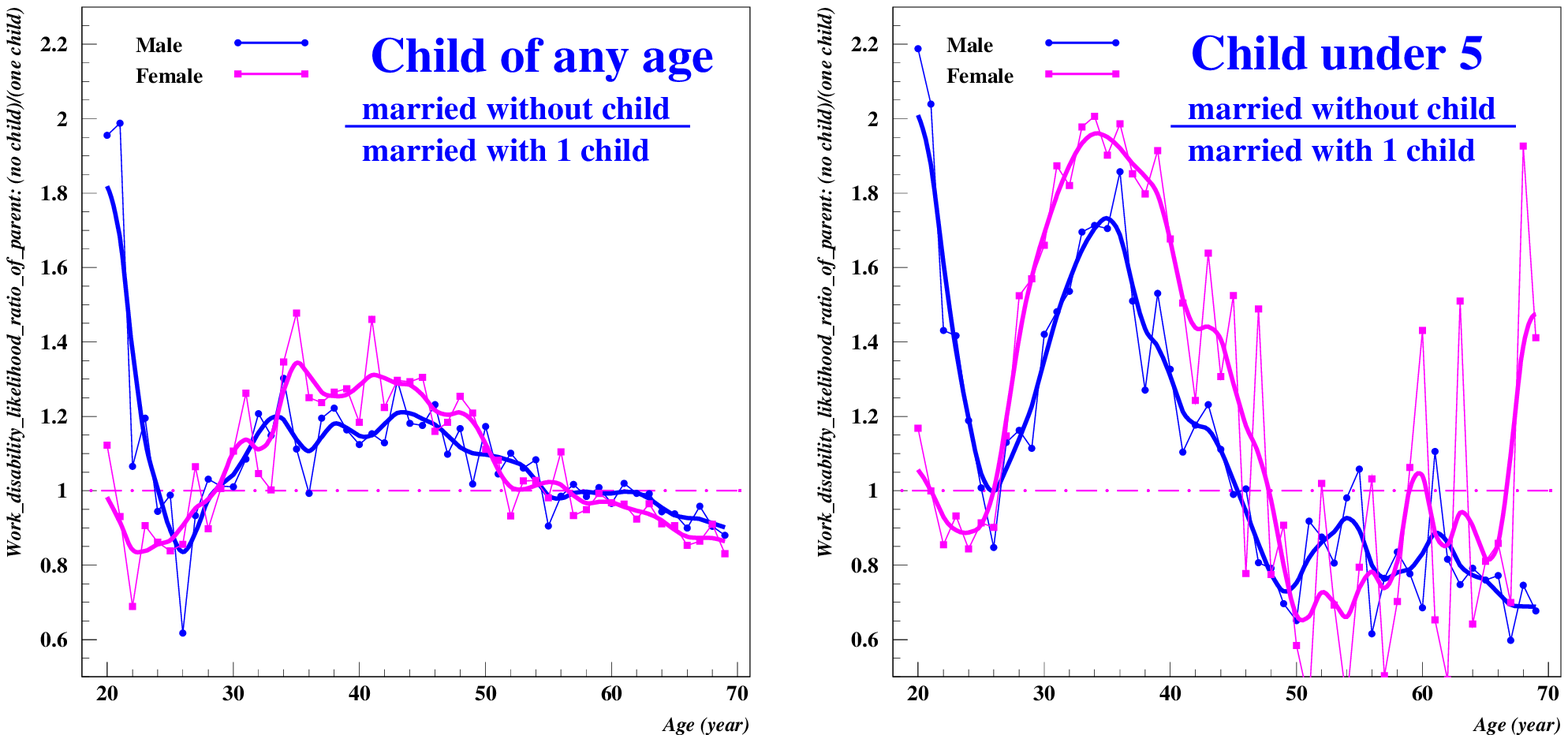}}
\qleg{Fig.\qhu 5a,b\qhv Changes in the disability ratio due
to the presence of a child (1990).}
{The fact that the curves are mostly above 1 shows that the
presence of a child reduces the disability.
{\bf (a) } All children are considered whatever their age. 
{\bf (b)} Only children less than 5 year old are considered.} 
{Source: 5\% sample of the US census of 1990 
downloaded from the IPUMS database (Ruggles et al. 2017).}
\end{figure}
%

Fig. 5a shows that for almost all ages until 60, the
disability is higher for couples without children
than for couples with one child. It is true that there is a small
interval between the ages of 20 and 28 where the disability
of mothers with a child is slightly higher than the disability of 
childless mothers. This may be due to
child-bearing health problems. It is not uncommon that
work must be discontinued in order to prevent a miscarriage
or a premature birth.

\qA{Comparison with the observations by Modig et al. (2017)}

In a very interesting paper Modig et al. (2017) give
estimates for the health benefits of having children for persons
aged 60 and over. Their study is based on the computerized
``Register of the Total Population'' (TPR) that exists in
Scandinavian countries%
\qfoot{For obvious confidentiality reasons this register is
not publicly available.}%
.
As summarized in graphical form in our conclusion
their results are quite compatible with ours.
It would be quite interesting to be able to compare
similar results for ages under 60 for which the 
disability and death ratios
are higher and therefore easier to compare.
We hope this will be possible in the future.
\qpar
A typical result of their study is the following.\qL
For men and women at age 60 (whether married or not)
the presence of (at least) one child resulted in a life expectancy
higher by 7.4\% than for childless men and women%
\qfoot{This percentage was obtained by dividing the life
expectancy benefit given in the paper
by the life expectancy at 60 in Sweden, namely 23.6 years in 2010.}%
.

\qA{Presence of a child under 5}

Fig. 5b analyzes the same situation but limited to
children under five. The most obvious observation
is that the peak values are notably higher for both
fathers and mothers, something which seems to agree
with our intuitive perception.\qL
Needless to say, children under 5 are
very uncommon for women over 45. They
may of course have adopted or step children but
such cases are rare. Consequently this
age interval is marked by large fluctuations. 
\qpar

The fact that the ``under age 5'' condition has
a significant impact encourages us to consider
more closely the effect of the age of the child.
It turns out that the IPUMS database has two variables
that allow us to know the child's age when there is only one.
Denoted YNGCH and ELDCH, these IPUMS variables give the ages of
the youngest and eldest child. When there is only one
child, YNGCH and ELDCH are equal and give the relevant age.
\qpar

In Fig. 6 the reference is the case of children of less
than one year of age (referred to as age 0). 
For the three parent age groups shown,
the disability ratio increases with child's age at least until 
the age of 10.

%
\begin{figure}[htb]
\centerline{\psfig{width=8cm,figure=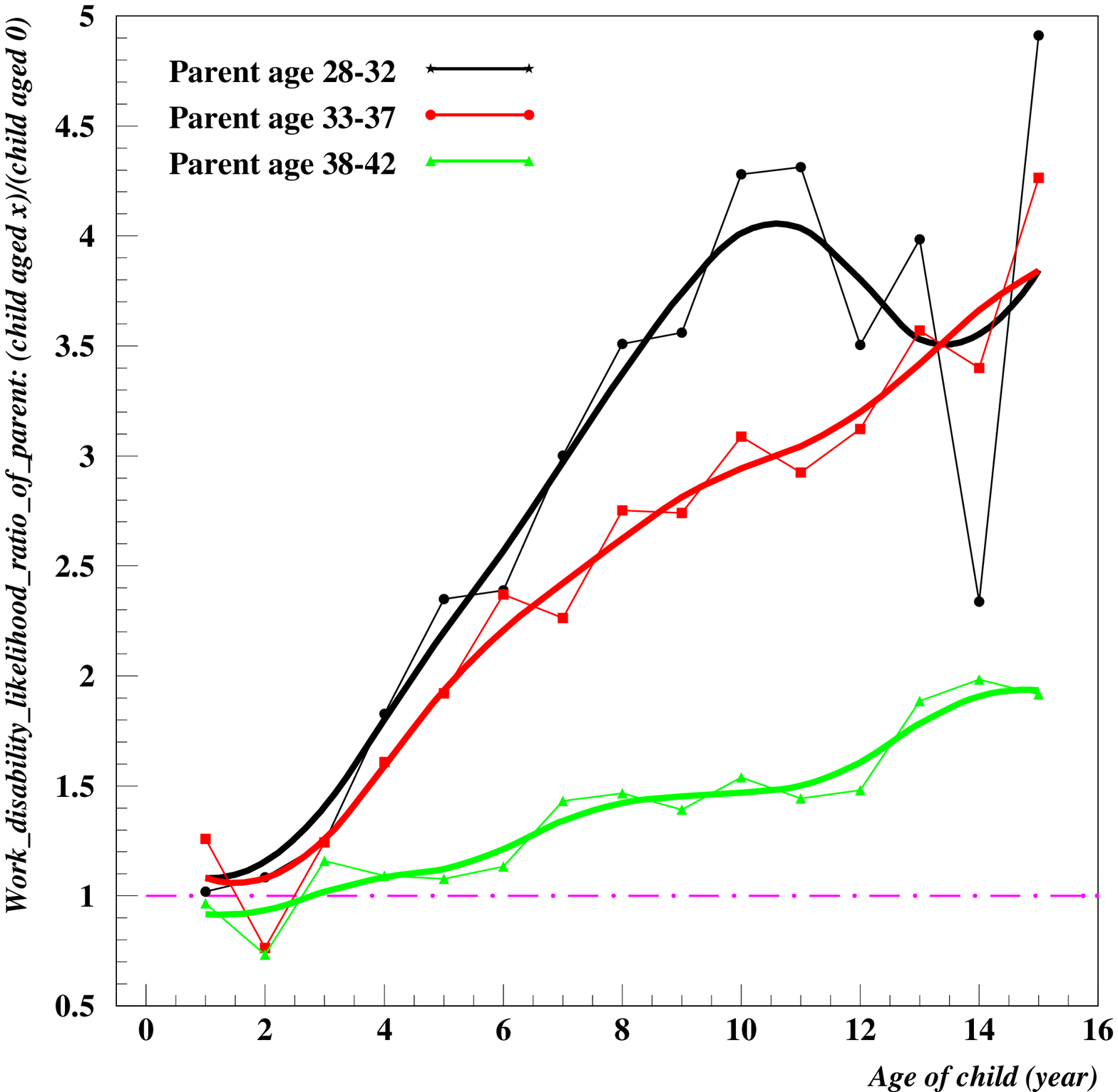}}
\qleg{Fig.\qhu 6\qhv Disability likelihood of parents 
according to the age $ x $ of their child;
the likelihood was normalized to make it
equal to 1 for $ x=0 $.}
{The fact that all curves are above 1 and going up shows that the
disability of the parents
is smallest with young babies and that it increases
with child age.} 
{Source: 5 \% sample of the US census of 1990 
downloaded from the IPUMS database (Ruggles et al. 2017).}
\end{figure}
%
\qI{Impact of an age gap between husband and wife}

%
\begin{figure}[htb]
\centerline{\psfig{width=8cm,figure=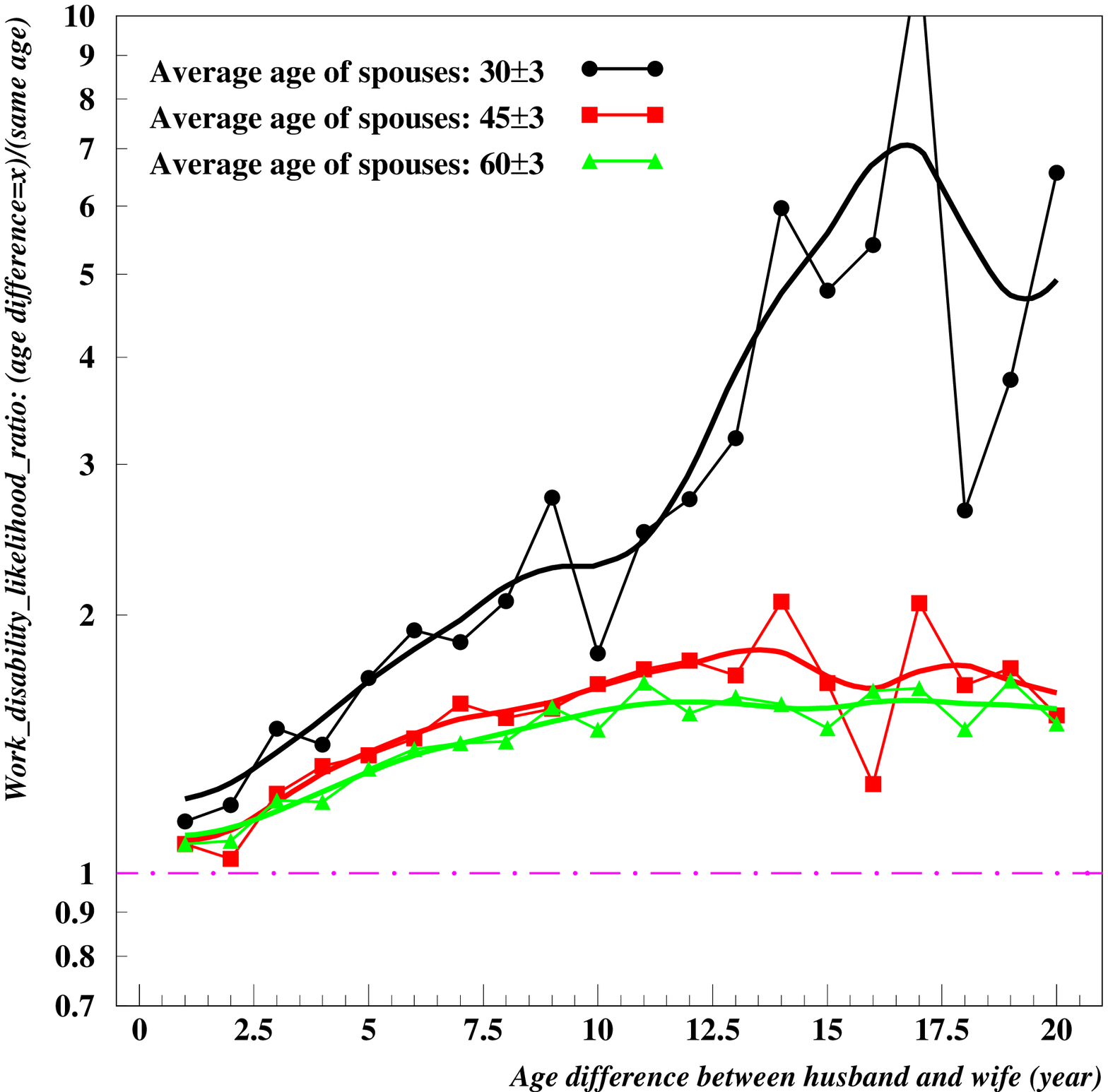}}
\qleg{Fig.\qhu 7\qhv Disability likelihood of spouses as a function
of their age difference $ x $; the likelihood
was normalized to make it equal to 1 for $ x=0 $.}
{The fact that all curves are above 1 and are upgoing shows that the
disability is smallest for $ x=0 $ and increases
with the age gap. } 
{Source: 5\% sample of the US census of 1990 
downloaded from the IPUMS database (Ruggles et al. 2017).}
\end{figure}

In Fig. 7 we investigate the influence of an age gap between
husband and wife. For that purpose the first requirement
is to be able to identify the pairs of record lines corresponding
to married couples.
This can be done thanks to the variable SERIAL
which is the identification number of each household.
The whole procedure can be summarized as follows.\qL
First, we select
the records for which the variable NUMPREC (number of
persons in the household) is equal to two. Most, but not
all of them, are of course married couples. 
The latter can be extracted thanks to the marital status variable.
In this way, one gets all married couples who are
without children. The children must of course be excluded
for otherwise they would affect the disability of the parents.
Finally, we select the couples with the SERIAL variable
and measure their disabilities. 
\qpar

Fig. 7 suggests that a non-zero age gap increases the disability ratio
particularly for young partners. Under our conjecture 
of an inverse connection between disability and interaction,
it means a reduction in interaction. More precisely,
one should say that there is a reduction in the health-linked
interaction for, as already observed, there can be other
forms of interaction that do not translate into health benefits.

\qI{Conclusion}

There can be two readings for the present paper. The first
and most
direct is in terms of disabilities and health. The second one
which is in terms of bond strength is based on what we called
Durkheim's conjecture. Here we will summarize our results
separately for the two interpretations.

\qA{Health effects seen in an evolutionary perspective}

Can 
the outcomes observed in the paper be viewed
in a unified manner? In other words can we divine a thread
which connects them?
The following 
characteristics seem to derive from
what may be called an evolutionary principle.
\qbu Married people have better health than unmarried
and this effect is at a maximum for young adults
between 20 and 30 (Fig.4b),
\qbu In addition such young couples have better health when
          husband and wife are of same age (Fig.7)
and their health is even better when they have
    already a young child (Fig. 5b).
\qpar

Clearly these successive characteristics all favor people 
who are in optimum condition to have children. In other
words, they may be viewed as important contributors to the
sustainability and optimization of the human reproduction
process.
\qpar

It is true that when one compares different species it does
not always make sense to suppose reproduction
optimization. For instance, some birds will have to fly 10,000 km
to their nesting area, a feat that will cause many losses,
while a closely related species may be able to 
reproduce without migrating at all. 
However, within one and the same species there must be a
form of optimization
at work in order to keep the species viable. 
This should be particularly
true for species whose reproduction margin is fairly narrow.

\qA{Basic questions about bond strength}

The results presented in this paper answer some basic questions
which we have had in mind for many years.
\qbu How does the strength of the interaction between parent
and children compare with the interaction between husband and wife?
\qbu How does the parent-child interaction change as the child
becomes a teenager? 
\qbu How does the strength of the
marital bond depend upon the partners' age gap?
\qpar

Needless to say, on all such questions
everybody may have an opinion based on personal
experience. The key-point was how
to address these questions in a {\it scientific way}.
The answers given in this paper were based on
Durkheim's long standing conjecture of a connection
between social isolation and suicide rates or more generally
death rates. By using the quasi-equivalence established
previously between death rates and work disability frequencies
we can find relative estimates for bond strengths.

%
\begin{figure}[htb]
\centerline{\psfig{width=16cm,figure=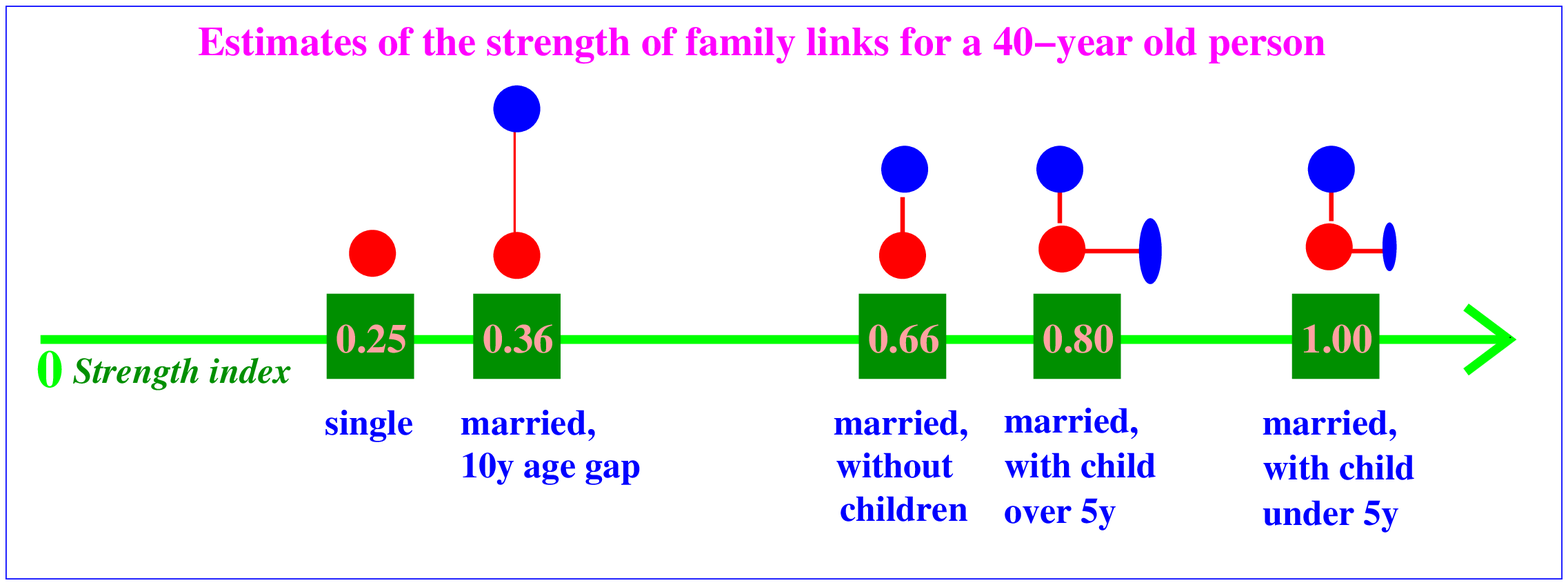}}
\qleg{Fig.\qhu 8a\qhv Summary of bond strengths $ b $ 
between 40-year old persons and their family members
in various situations.}
{The estimates for $ b $ 
were derived from the disability likelihood
$ d $ read in Fig. 1b, 2, 4 and 6, through
the relationship $ b\sim 1/d $. The situation
``married with a child under 5'' was taken as
the reference level and assigned the value $ 1.00 $.} 
{Sources:  Fig. 1b, 2, 4 and 6}
\end{figure}
%
%
\begin{figure}[htb]
\centerline{\psfig{width=16cm,figure=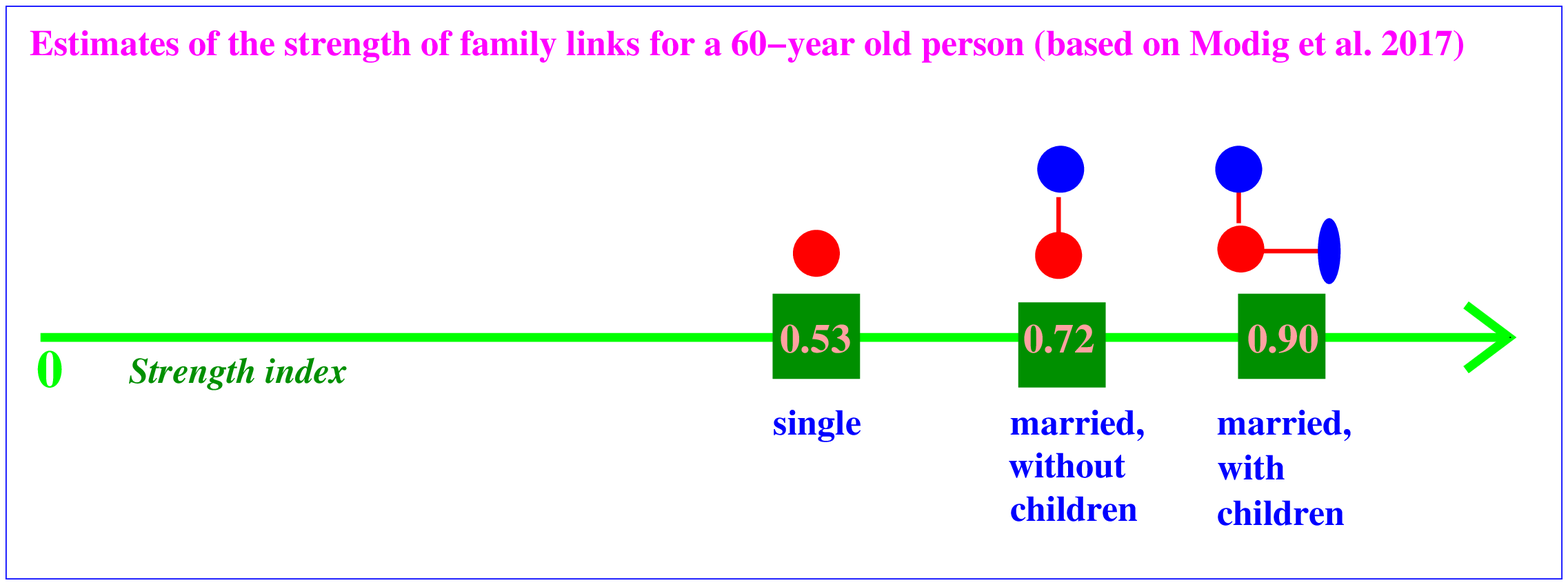}}
\qleg{Fig.\qhu 8b\qhv Summary of bond strengths $ b $ 
between 60-year old persons and their family members
in various situations.}
{The estimates for $ b $ 
were derived from the death rates computed in the study by
Modig et al. (2017) in the same way as in Fig. 7a.
The situation
``married with children'' was taken as the reference level
and the value $ 0.9 $ was attributed by comparison with
Fig. 7a.} 
{Source:  Modig et al. (2017) and complementary data from the
same study (personal communication).}
\end{figure}

These estimates seem
consistent with one another.

\qA{Other perspectives}

Before closing this paper, it must be mentioned that
Durkheim's perspective based on interaction strength is not the
only one that has been considered. As a matter of fact,
nowadays the dominant perspective seems to be one
based on {\it individual} psychological characteristics.
As typical of this perspective  one can
mention a paper by Walter Gove (1973). In a sense this
conception returns to the anthropomorphic
conception prevailing before Durkheim. Its main drawback
is that it is difficult to test it in any meaningful way for
it incorporates a variety of {\it ad hoc}
psychological explanations. 
\qpar
Let us illustrate this point by an
example from Gove (1973). Observing in his Table 8 that for 
death by tuberculosis widowed males in the age-group 25-34
(and in the time interval 1959-1961) have a death rate 11.5 higher
than the rate of married males of same age, the author
suggests (p. 47) that
being ``more aggressive and willing to take risk'',
men would be ``less apt to enter treatment and persist in
the prolonged and careful treatment that is necessary
for tuberculosis''.
From a scientific perspective
recourse to specific arguments (in this case
special features of tuberculosis treatment) 
for different causes of
death is hardly satisfactory in so far as 
the high death rate ratio
for young males is by no means limited to tuberculosis.
Only
testable {\it predictions} would be convincing.

\qA{Prospects}

What should be done next?\qL
In physics there are several ways to measure inter-molecular
interactions. One is to measure the amount of energy
necessary to break the molecular bonds. 
For a liquid, simply heating it to its boiling point does
this.
Another approach is to estimate
the parameters of the Lennard-Jones potential
on the one hand and the average inter-molecular distance on the
other hand. It is the fact that all such methods lead
to consistent results that
makes the methodology truly convincing. 
\qpar
Ideally we should follow the same path here. 
We have already indicated that in a family
the health-related interaction is probably not the only form
of interaction. Finding other interaction indicators
would be enlightening.

\appendix

\qI{Appendix A. Data selection and analysis}

As an example, we wish to compare non-married persons who live alone
to married couples without children. 

\qA{Data selection and extraction from the IPUMS data base}

On the IPUMS database (Ruggles et al. 2017)
there are two kinds of variables (i) those which refer
to the household and (ii) those which refer to individuals.
Here we will use one household variable, namely the
number of persons who compose the household 
(noted NUMPREC in the IPUMS coding system)
and the following 5 variables pertaining to individuals.
\qbu The number of children of each individual (NCHILD)
\qbu The sex and age of each individual (SEX, AGE)
\qbu The marital status of each individual: noted
MARST it offers 6 options but here
we will use only 2, namely 
1=married (spouse present), 6=never married. 
\qbu A variable (noted DISABWRK) which describes the
disability status of the person. IPUMS offers
4 options but here we will use only two, namely:
1=No disability affecting work, 3=Disability prevents work.
\qpar

Once these 6 variables have been selected, specific limits
will be chosen within IPUMS for 3 of them. 
\qbu Only persons without children (NCHILD=0) will be retained.
This eliminates all couples with children
and markedly reduces the size of the file.
\qbu As the disability variable is defined only over
the age-range (16,89) we will limit our age interval to (20,89).
\qbu Moreover, as we consider only two marital
situations. we eliminate all records for which MARST is not
equal to 1 or 6.
\qpar

Finally we select the 5\% sample of the census of 1990;
remember that the work disability variable exists
only for the censuses of 1980 and 1990.
\qpar

This selection procedure yields
a file of 3,150,056 lines (31 M once decompressed) 
where each line corresponds to one person.
By way of illustration, the first data lines reads:\qL
$ \rightarrow $ 050102361. \qL
If one introduces spaces between the 6 successive variables, 
one gets a more readable data line: \qL
$ \rightarrow $ 05 0 1 023 6 1. \qL
These variables have the following meaning:\qL
$ \rightarrow $ {\small
05 persons in household,  0 child, 1=male, 023=age,
  6=single, 1=no disability.}\qL
On should make a distinction between 05, 0, 023 which are
``real'' data and 1, 6, 1 which are code numbers that
take significance only through the conversion tables
given in IPUMS.
\qpar

\qA{Analysis of the data}

The file is read line by line and all
non-married persons who live {\it alone}
(i.e. NUMPREC=1 and MARST=6) are collected.
Let $ HZ $ denote
the number of all persons who correspond to this specification
Among them, a number $ HZd $
have a disability (i.e. DISABWRK=3). Thus, non-married
persons will be disabled with a probability $ PZ=HZd/HZ $.
\qpar

In the same way, we collect all couples who live without
any other person in the same household by selecting the lines with  
NUMPREC=2 and MARST=1; a fraction $ P1=H1d/H1 $ of them
will have a disability.


\qI{Appendix B: Confidence intervals vs. phenomenon variability}

In this paper we are interested in stable {\it structural}
properties. A relationship between marital status and 
disability which would be valid for only one (or a few)
years would be of little interest. In other words, beyond the 
purely statistical notion of confidence intervals 
there is the question of the variability of the
phenomenon under consideration.
In our perspective the later is of greater significance
than the former.
Below these two aspects are discussed successively.

%
\begin{figure}[htb]
\centerline{\psfig{width=16cm,figure=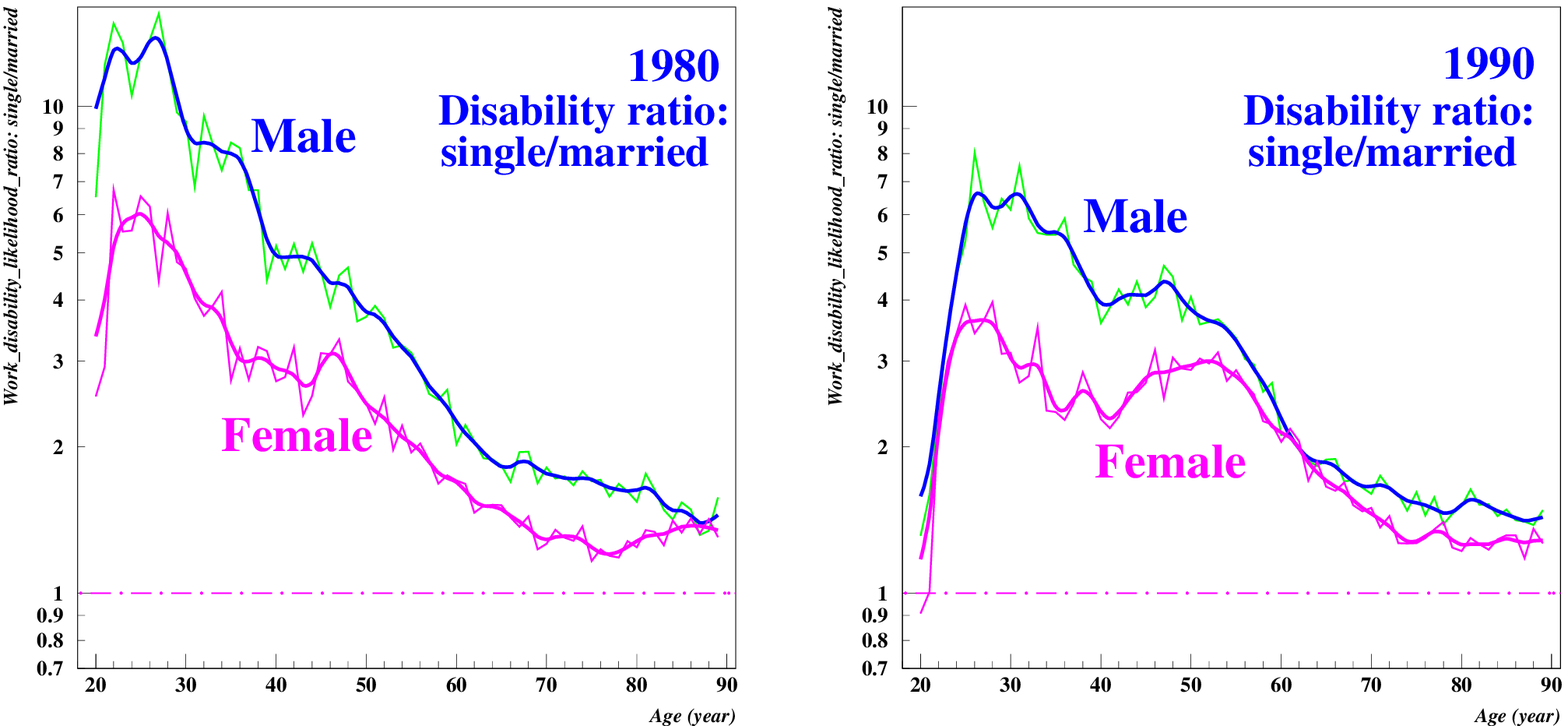}}
\qleg{Fig.\qhu B1a,b\qhv Variability assessment:
single/married disability ratio in 1980
and 1990.}
{Here ``single'' means that the persons are not living with
other persons in the same household; ``married'' means without
children and without anyone else in the same household.
The fact that all curves are above 1 is consistent
with the result already known from death rate data.  
{\bf (a)} The 1980 sample consisting of single persons
and married couples without children comprises 3,150,056
individuals.
{\bf (b)} In 1990 it comprises 3,695,444 persons.} 
{Source: US censuses of 1980 and 1990 from 
the IPUMS data base (Ruggles et al. 2017).}
\end{figure}
%

\qA{Error bars}

In Richmond et al. (2017) error bars (or rather error bands)
showing confidence intervals for every age
were drawn. However, 
such error bands can impair the readability of the graphs.
Here we have no need for accurate error bars for we are mostly
interested in the overall shape of the disability curves.
That is why we will use an alternative representation
taking advantage of the fact that
the magnitude of the confidence intervals is well
represented by the amplitude of yearly fluctuations.
These fluctuations 
around the 5-year centered moving average
define the uncertainty margin almost quite as well as 
the error bands used previously. 

\qA{Variability}

With respect to error bars, the comparison 
between Fig. B1a and B1b
tells us something else of interest. It shows that from
one decennial census to the next, the disability ratio changes
far more than what the confidence interval within one and the
same sample would suggest. This is of course hardly surprising
for in an interval of 10 years, many conditions may change which
may to some extent affect the disability ratio.
This shows that confidence intervals within
a given census are in fact of limited interest for the 
purpose of assessing the variability of the phenomenon
under consideration.


%
\count101=0  \ifnum\count101=1

\qI{Effect of the presence of other household members}

When a distinction is made according to
marital status, mortality statistics
does not specify whether or not other
persons are present in the household. The most obvious case
is of course the presence of children in households of
married couples, but it may also be young non-married
persons still living in the same household as their parents.
\qpar
In contrast, in Fig, 1a,b we considered single and married
persons in a strict sense that is to say without any other
person present in the household. We will introduce the notation
single/married to refer to the strict definition
and the new notation single+/married+ when we refer to the broad
meaning with possibly other persons present.
\qpar

How is
Fig. 1 changed by shifting to single+/married+.
Actually, it is better to analyze this transition
in two steps, namely: 
single/married $ \rightarrow $ single+/married on the one hand and 
single/married $ \rightarrow $ single/married+ on the other hand.
These transitions are shown in Fig.3a,b. Not surprisingly,
the most important change comes from the presence of children
in the households of married couples.
Finally, the transition  
single/married $ \rightarrow $ single+/married+ is the composition
of the two previous ones. 
\qpar
%
\begin{figure}[htb]
\centerline{\psfig{width=16cm,figure=sm.eps}}
\qleg{Fig.\qhu C1a,b\qhv Disability ratio
single(+)/married(+) with or without a presence of other persons.}
{The fact that all curves are above 1 shows that,
as already known from death rate data,
disability is smaller for married people
than for never-married persons. The present graphs give a more
detailed description.
{\bf (a)} Transition from single/married (without
any one else) to the case single+/married where other persons are
allowed in the household of the single person. 
It can be seen that there is a substantial difference only
for the ages around 20.
{\bf (b)}  Transition from single/married (without
any one else) to the case single/married+ where other persons are
allowed in the household of the married couple.
The peak becomes broader but, rather surprisingly, not higher.
The curve for the case single+/married+ is close to single/married+;
it was not shown here for the sake of clarity.
In these graphs, for the sake of clarity, the fluctuations were not
shown because they are very similar to what can be seen in Fig. 1 and
Fig.2.} 
{Source: US census of 1990 from the IPUMS
data base (Ruggles et al. 2017).}
\end{figure}
%
\fi

\vskip 10mm

{\bf \large References}

\qparr
Bertillon (L.-A.) 1872:
Article ``Mariage'', in: Dictionnaire Encyclop\'edique des Sciences
M\'edicales, [Encyclopedic Dictionary of the Medical Sciences],
vol. 5, 2nd series. p. 7–52. \qL
[Available on ``Gallica'', the website of digitized publications of
the French National Library]

\qparr
Bertillon (L.-A.) 1879: Article ``France'', 
in: Dictionnaire Encyclop\'edique des Sciences M\'edicales, 
[Encyclopedic Dictionary of the Medical Sciences], 
vol. 5, 4th series, p. 403-584. \qL
[Available on ``Gallica'', the website of digitized publications of
the French National Library.]

\qparr
Christenson (K.), Vaupel (J.W.) 2011: Genetic factors
in adult mortality. Pages 399-410
in: Rogers (R.G.), Crimmins (E.M.) editors: International
handbook of adult mortality. Springer, Heidelberg.

\qparr
Durkheim (E.) 1897: Le suicide. Etude de sociologie.
F. Alcan, Paris.\qL
[A recent English translation is: ``On Suicide'' (2006), Penguin
Books, London].

\qparr
Gove (W.R.) 1973: Sex, marital status and mortality.
American Journal of Sociology 79,1,45-67.\qL
[This study relies on the data given in ``US Public Health
Service'' 1970 cited below.]

\qparr
Modig (K.), Talb\"ack (M.), Torssander (J.), Ahlbom (A.) 2017:
Payback time? Influence of having children on mortality in
old age.
Journal of Epidemiology and Community Health 71,424-430.

\qparr
Osler (M.), McGue (M.), Lund (R.), Christensen (K.)
2008: Marital status and twins' health and behavior: 
an analysis of middle-aged Danish twins.
Psychosomatic Medicine 70,4,482-487. 

\qparr
Richmond (P.), Roehner (B.M.) 2016a: 
Effect of marital status on death rates. Part 1: High accuracy
exploration of the Farr-Bertillon effect.
Physica A, 450,748-767.

\qparr
Richmond (P.), Roehner (B.M.) 2016b: 
Effect of marital status on death rates. Part 2: Transient
mortality spikes.
Physica A, 450,768-784.

\qparr
Richmond (P.), Roehner (B.M.) 2017: Impact of marital
status on health. To appear in Physica A.
Preprint available on the arXiv website
(arXiv:1704.00752, 30 March 2017).

\qparr
Ringb\'ack Weitoft (G.), Burstr\"om (B.), Ros\'en (M.) 2004:
Premature mortality among lone fathers and childless men.
Social Science and Medicine 59,1449-1459.

\qparr
Roehner (B.M.) 2007: Driving forces in physical, biological
and socio-economic phenomena. A network science investigation
of social bonds and interactions. Cambridge University Press,
Cambridge.

\qparr
Ruggles (S.), Genadek (K.), Goeken (R.), 
Grover (J.), Sobek (M.) 2017:
Integrated Public Use Microdata Series (IPUMS). 
University of Minnesota, Minneapolis (Minnesota).

\qparr
Santerre (R.E.), Neun (S.P.) 2013: Health economics.
Theory, insight, and industry studies. 6th edition.
South Western Cengage Learning.

\qparr
US Public Health Service 1970: Mortality from selected causes by
marital status. United States, Part A.
Rockville (Maryland).

\qparr
Wang (L.), Xu (Y.), Di (Z.), Roehner (B.M.) 2016:
Effect of isolation on life expectancy of red imported
fire ants {\it Solenopsis invicta} and tephridid fruit fly
{\it Bactrocera dorsalis}.
Acta Ecologica Sinica 36,252-255.

\end{document}